\documentclass{article}
\usepackage{spconf,amsmath,graphicx,multirow,amsfonts, hyperref, amssymb, xcolor, float}

\title{Audio Match Cutting: Finding and Creating Matching \\ Audio Transitions in Movies and Videos}
\name{Dennis Fedorishin$^{1,2\star}$\thanks{$^\star$Done during internship at Dolby Laboratories}, Lie Lu$^1$, Srirangaraj Setlur$^2$, Venu Govindaraju$^2$}
\address{$^1$Dolby Laboratories, $^2$University at Buffalo}
\begin{document}
\ninept
\maketitle
\begin{abstract}

A ``match cut" is a common video editing technique where a pair of shots that have a similar composition transition fluidly from one to another. Although match cuts are often \textit{visual}, certain match cuts involve the fluid transition of \textit{audio}, where sounds from different sources merge into one indistinguishable transition between two shots. In this paper, we explore the ability to automatically find and create ``audio match cuts" within videos and movies. We create a self-supervised audio representation for audio match cutting and develop a coarse-to-fine audio match pipeline that recommends matching shots and creates the blended audio. We further annotate a dataset for the proposed audio match cut task and compare the ability of multiple audio representations to find audio match cut candidates. Finally, we evaluate multiple methods to blend two matching audio candidates with the goal of creating a smooth transition. Project page and examples are available at: \url{https://denfed.github.io/audiomatchcut/}


\end{abstract}


\vspace{-0.5em}
\section{Introduction}
\label{sec:intro}

\begin{figure}[t]
  \centering
  \includegraphics[width=0.8\columnwidth]{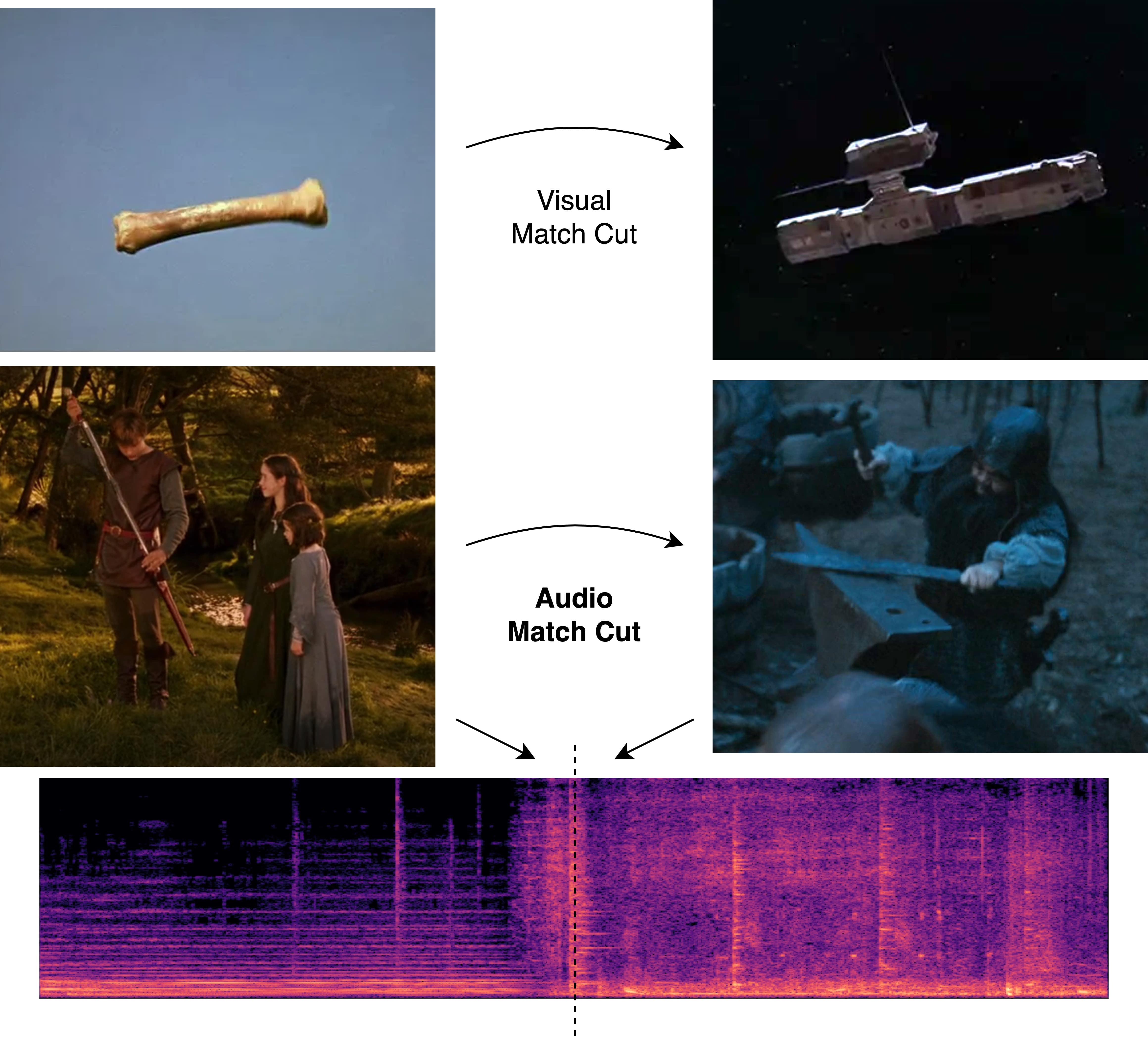}
  \vspace{-1em}
  \caption{Example match cuts in movies. In \textit{2001: A Space Odyssey} \cite{spaceodyssey} (top), two different visuals transition fluidly based on the similar size and shape of the objects. In \textit{The Chronicles of Narnia: The Lion, the Witch and the Wardrobe} \cite{narnia} (bottom), The sound of a sword clinking within its sheath matched to the strike of a hammer in the next scene, creating a seamless audio match across scenes.}
  \label{fig:concept}
\vspace{-1.5em}
\end{figure}

In movies and videos, the ``cut" is a foundational editing technique that is used to transition from one scene or shot to the next \cite{cutting2016evolution}. The precise use of cuts often crafts the story being portrayed, whether it controls pacing, highlights emotions, or connects disparate scenes into a cohesive story \cite{kozlovic2007anatomy}. There are many variations of cuts that are used across the film industry, including smash cuts, reaction cuts, J-cuts, L-cuts, and others. One specific cut is the ``match cut", which is a transition between a pair of shots that uses similar framing, composition, or action to fluidly bring the viewer from one scene to the next. Match cuts often match \textit{visuals} across each scene, either through similar objects and their placement, colors, or camera movements \cite{douglass1996art}. However, match cuts can also match \textit{sound} across scenes, where sound between two scenes transition seamlessly between each other. These audio match cuts (also referred to as ``sound bridges") either blend together sounds or carry similar sound across scenes, often from different sound sources, to create a fluid audio transition between them \cite{thompson2013grammar}. Figure \ref{fig:concept} shows examples of visual and audio match cuts found in movies.







Along with cutting, video editing as a whole is a time-consuming process that often involves a team of expert editors to create high-quality videos and movies. When performing tasks like match cuts, it often involves a manual search across a collection of recorded content to find strong candidates to transition to, which becomes a time-consuming and tedious manual process \cite{matchcut}. As a result, AI-assisted video editing has emerged as a promising area of research, with the goal of aiding editors improve the speed and quality of editing. Recent works focus on improving the understanding of movies, from detecting events and objects within them, like speakers \cite{irie2010automatic}, video attributes like shot angles, sequences and locations \cite{ave}, and understanding various cuts \cite{moviecuts}. Beyond understanding videos, full editing tools have been proposed including shot sequence ordering \cite{ave}, automatic scene cutting \cite{pardo2021learning}, trailer generation \cite{irie2010automatic}, video transition creation \cite{shen2022autotransition}, and audio beat matching \cite{pei2023automatch}. Recently, \cite{matchcut} proposed a framework to automatically find frame and motion match cuts in movies. \cite{matchcut} collects a large-scale dataset of match cuts found in movies and further trains a classification network to retrieve match cut candidates, aiding video editors in finding and creating these match cuts. However, \cite{matchcut} only focuses on \textit{visual} match cuts. In our work, we expand upon this area and focus on the ability to automatically find create \textit{audio} match cuts.

\begin{figure*}
  \centering
  \includegraphics[width=0.98\textwidth]{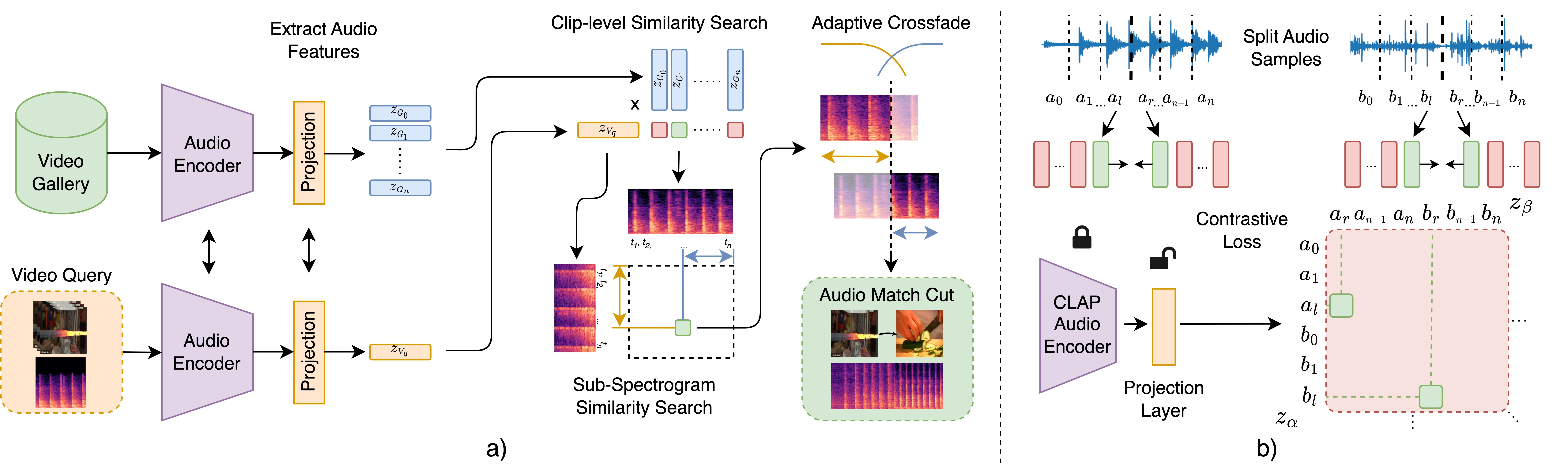}
  \vspace{-1.5em}
  \caption{a) Proposed Framework. Given a query video, we retrieve an audio match cut candidate from a video gallery and find the optimal transition point using a sub-spectrogram similarity search. Using the variance of the created similarity matrix, we adaptively select the crossfade length to blend both the query and match audio into a fluid audio match cut. b) Proposed ``Split-and-Contrast" contrastive objective. Each audio sample is split at a randomly selected frame, then the adjacent frames of the split are contrasted towards each other.}
  \label{fig:model}
\vspace{-1.5em}
\end{figure*}

At its core, creating audio match cuts involves retrieving candidate audio clips that are able to create high-quality match cuts. Retrieving similar audio samples have been explored in the music domain with music information retrieval systems that retrieve full songs based on small song snippets, using signal processing techniques \cite{six2014panako, fenet2011scalable} and more recently deep learning \cite{banerjee2022wav2tok, chang2021neural}. Similarly, performing audio transitions in the music domain is an often-used technique, both in music mixing and live DJ performances. In literature, both signal processing \cite{vande2018raw} and deep-learning-based \cite{chen2022automatic} techniques have been introduced to automatically create these transitions. However, finding and creating matching audio transitions has been unexplored in the context of movies and videos across a diverse set of sounds beyond music.

In this paper, we explore this problem and propose a self-supervised retrieval-and-transition framework, shown in Figure \ref{fig:model}, to automatically find and create high-quality audio match cuts. Our contributions in this paper are as follows:

\begin{itemize}
\item We introduce the problem of automatic audio match cut generation across diverse sounds and create two datasets for evaluating automatic audio match cutting methods.

\item We propose a framework where a coarse-to-fine audio retrieval pipeline first recommends matched clips, then a fine-grained transition method creates audio match cuts that outperform multiple baselines.

\end{itemize}

\vspace{-1.5em}
\section{Method}
\label{sec:method}

\vspace{-0.5em}

\subsection{Problem Definition}
\label{ssec:problemdef}


To model the real-world task of creating audio match cuts, we formulate our proposed audio match cut problem as a unimodal audio retrieval task. Specifically, given a query video clip $V_{q}$ and a collection of $n$ other video clips $G = \{i_1, i_2, ..., i_n\}$, the goal is to retrieve a video clip $G_i$ and create an audio transition such that $V_q \Rightarrow G_i$ creates a high-quality audio match cut. We formulate the retrieval as a maximum inner-product search (MIPS) between extracted $L^2$ normalized feature representations of the query and a gallery of the audio of video clips, $z_{V_q}, z_{G_i} \in \mathbb{Z}^d$, denoted by $z^*_{G_i} = \text{argmax}_i (z_{V_q}^T z_{G_i})$. After retrieving the top-k highest-similar gallery clips $\{G^*_{i}\}_{i=1}^k$, we perform a processing operation $f_p$ to blend the query and retrieved clips to create the final audio match cuts $\{f_p(V_q, G^*_i)\}_{i=1}^k$, where a user selects which match cuts to use out of $k$ recommendations.


\vspace{-0.5em}

\subsection{Data Collection}
\label{ssec:datacollection}

As the audio match cut problem is unexplored, we developed evaluation sets based on subsets of publicly available datasets, Audioset \cite{audioset} and Movieclips\footnote{Movieclips is collected from the \href{https://www.youtube.com/c/MOVIECLIPS/videos}{Movieclips YouTube Channel}}, to evaluate audio match cut generation methods. Audioset contains user-generated videos from YouTube and Movieclips contains high-quality movie snippets. For each dataset, we split each video into 1-second non-overlapping image-audio pairs where the image is the middle frame of the respective second of video, resulting in over 2M Audioset and 800k Movieclips samples. We selected to perform retrieval over 1-second pairs to balance between granularity and search complexity.


Next, we collect a query set of samples of a variety of natural sounds and sound effects, including sounds like engines revving, impulsive sounds like a hammer striking, doorbells, campfires, and other unique sounds seen in videos and movies. For each query, we label a set of match candidates based on two criteria that constitute a positive audio match: i) the pair must sound plausible if the audio is swapped between the query and match images. ii) the pair must sound perceptually similar in terms of pitch, rhythm, timbre, etc.

As labeling random pairs across all samples results in an unfeasible search space (over 4 trillion pairs), we use existing audio representations to help generate candidate audio matches. We hypothesize that since the main characteristic of audio match cuts is that the audio of both scenes are perceptually-similar, widely-available audio representations may be used as they often are trained with the goal of similar audio samples having high similarity. We use two simple representations, the MFCC and Mel-Spectrogram, and two deep representations, the audio encoders from CLAP \cite{clap} and ImageBind \cite{imagebind}. For MFCC and Mel-Spectrogram, we use a window of 2048 samples and hop length of 1024 samples. We flatten both representations along the time steps and use the resulting feature vectors for retrieval. For CLAP \cite{clap} and ImageBind \cite{imagebind}, we use their respective spectrogram generation parameters and use the resulting audio encoder feature vectors for retrieval. We use the MIPS operation described in Section \ref{ssec:problemdef} to create the audio match cut candidates for labeling. All audio used in this work is sampled at 48kHz.

Since we used audio representations to collect audio match candidate pairs and label only those pairs, our evaluation set tends to favor the highest-similar candidates of each representation. To address this bias and create a more comprehensive evaluation, we randomly sample 100 negative matches for each query in the Audioset and Movieclips dataset. By randomly selecting samples out of millions of 1-second samples, there is a very unlikely chance that these samples in fact are positive audio matches. The resulting Audioset evaluation set has a gallery of 12,350 labeled samples spread across 102 queries, and the Movieclips evaluation set has a gallery of 8,289 labeled samples spread across 66 queries. Each query has an average of 123 labeled samples and 10 positive matches.


\vspace{-0.5em}

\subsection{Audio Match Cut Representation Learning}
\label{ssec:retrieval}



In Section \ref{ssec:datacollection}, we utilize existing audio representations to generate audio match cut candidates. However, existing audio representations are not directly aligned for the audio match cut task, which aims to retrieve perceptually-similar audio from different scenes, differing from existing retrieval tasks. As a result, existing audio representations may produce sub-optimal audio match cut candidates.


Lacking labeled data for audio match cutting, we propose a self-supervised learning objective to create an audio representation that effectively retrieves high-quality audio match cut candidates. Our objective leverages already-edited videos based on the notion that given a query audio frame of a video, an audio frame that results in a high-quality match cut is the next successive frame in the same video, as the entire video has been previously edited to have continuous audio. We model this characteristic as ``Split-and-Contrast", shown in Figure \ref{fig:model}b, where adjacent audio frames in two splits of a video are trained to have high similarity, while contrasting away other non-adjacent audio frames.

Given a batch of $N$ audio samples that have $n$ audio frames, for every sample, we extract a feature representation $z$ from each frame, $\{z_k\}_{k=1}^{n} \in \mathbb{Z}^{d \times n}$, where $d$ is the feature representation size. We then randomly select an index to split the $N$ sets of features into left/right sections $z_\alpha \in \mathbb{Z}^{d \times n_\alpha}$ and $z_\beta \in \mathbb{Z}^{d \times n_\beta}$, of length $n_\alpha$ and $n_\beta$, such that $n_\alpha + n_\beta = n$. For each left/right section in $N$, we denote the adjacent frames as $z_{k_l} = z_{\alpha_{n_\alpha}}$ and $z_{k_r} = z_{\beta_0}$, corresponding to the last frame in the left section, and first frame in the right section, respectively. Then, we define a contrastive learning formulation for a batch of $N$ samples to learn a representation that produces high similarity for only the adjacent frames in the split sections, and low similarity for all other pairs: 

\vspace{-1em}

\begin{equation}
\mathcal{L}_{S\&C}(N) = - \text{log} \left( \frac{\sum_{k=1}^{N}\text{exp}(z_{k_l}^T z_{k_r} / \tau)}
{\sum_{i=1}^{N\cdot n_\alpha}\sum_{j=1}^{N\cdot n_\beta}\text{exp}(z_{\alpha_i}^T z_{\beta_j} / \tau)} \right)
\end{equation}
\vspace{-0.5em}

$z_{a}^T z_{b}$ denotes the inner product of $L^2$ normalized vectors, and $\tau$ denotes a temperature parameter for softmax. This formulation is similar to InfoNCE \cite{oord2018representation} and N-Pair \cite{sohn2016improved} loss, modified to allow multiple positives in a single loss computation. By maximizing the similarity of two adjacent frames in a split audio sample, we expect the model to learn to retrieve perceptually-similar audio frames that result in high-quality transitions. We use the pretrained CLAP \cite{clap} audio encoder, based on the HTSAT \cite{htsat} architecture, and the CLAP \cite{clap} linear projection layers. We also use the spectrogram creation and preprocessing steps defined in \cite{clap}, using an audio frame size of 1-second. We found fine-tuning the CLAP \cite{clap} projection layers with a frozen encoder works better than end-to-end finetuning, suggesting that ``Split-and-Contrast" is better suited for aligning existing audio feature representations for the audio match cut task.



We train the projection layers using 200k random Audioset samples for 20 epochs using the Adam \cite{adam} optimizer, learning rate of $10^{-4}$, batch size of 2048, and temperature $\tau$ of 0.1. Each sample has ten 1-second audio frames, such that $n_\alpha + n_\beta = 10$.



\vspace{-0.5em}

\subsection{Audio Transition}
\label{ssec:transition}

One common method of transitioning between two audio samples is the crossfade, where the first audio clip fades out while simultaneously fading the second clip in, resulting in a smooth transition \cite{archibaldcross}. However, creating high-quality transitions using crossfade often requires manual tuning of the crossfade length, based on the audio characteristics \cite{lupsa2020audio}. In this section, we describe our audio transition method that improves upon simple crossfade by i) first finding a specific transition point within the 1-second clip, and ii) adaptively selecting a more optimal crossfade length based on the audio characteristics. 

\begin{figure}[t]
  \centering
  \includegraphics[width=\columnwidth]{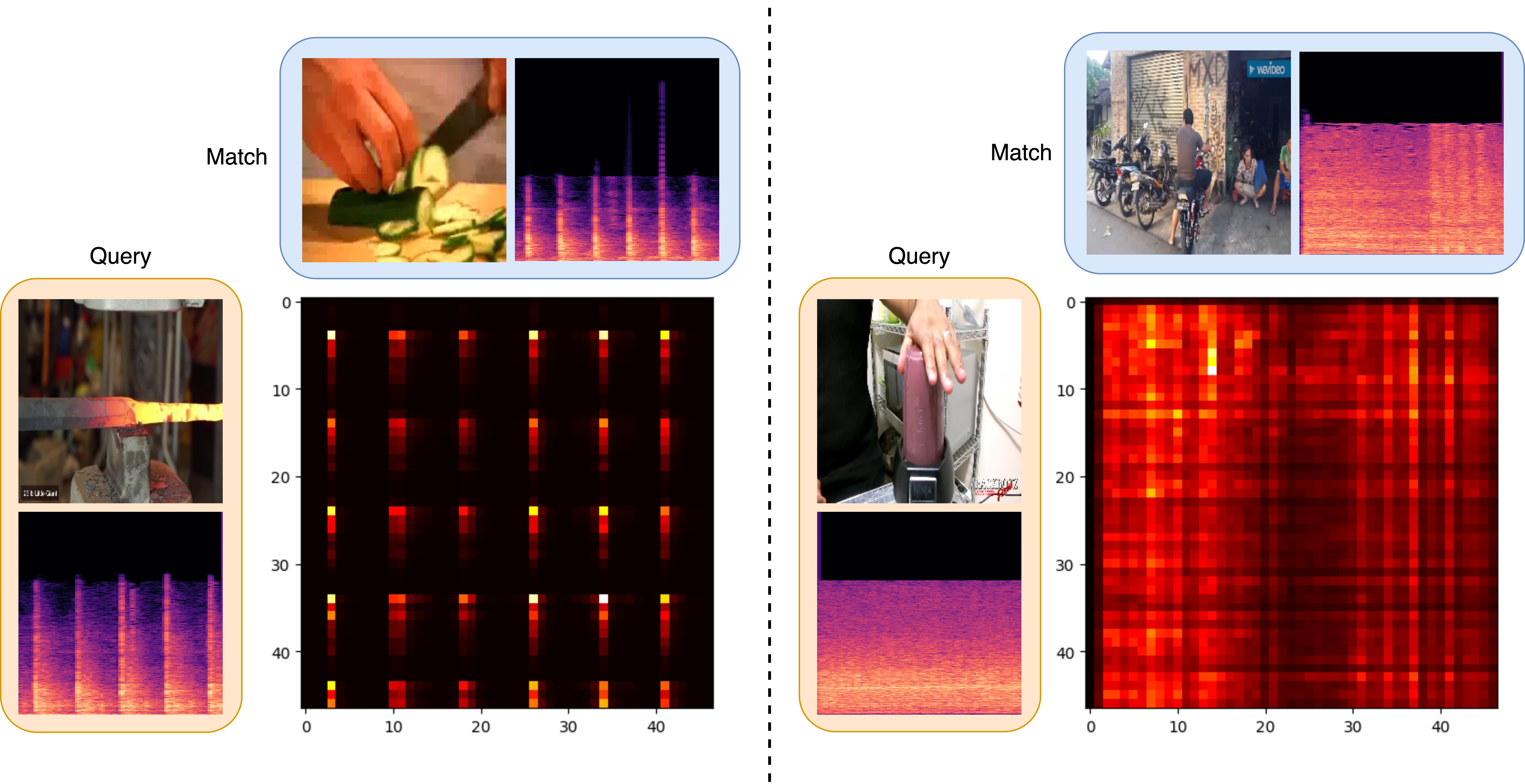}
  \vspace{-2em}
  \caption{Example sub-spectrogram similarities of audio match cuts: A forging hammer striking matched with a knife chopping (left) exhibits high similarity on each strike occurrence. A blender matched with a motorcycle revving (right) shows a smoother similarity matrix, allowing for plausible transitions across multiple time steps.}
  \label{fig:examplesim}
\vspace{-1.5em}
\end{figure}



Since we perform retrieval using 1-second audio clips, the matched clips may be overall strong candidates for an audio match cut, but the exact borders of each still may not align well for a direct transition. Therefore, we propose an operation to find an optimal transition point between the query and matched clip at the spectrogram time-step-level, named ``Max Sub-Spectrogram" similarity search, shown in Figure \ref{fig:model}a. 

Given a Mel-spectrogram representation of the query audio $S_Q\in\mathbb{R}^{f\times t}$ and matched audio $S_M\in\mathbb{R}^{f \times t}$, where $f$ and $t$ denote the frequency bins and time steps, respectively, we calculate the inner product of two spectrograms across time steps, yielding a similarity matrix $M = S_Q^T S_M \in\mathbb{R}^{t \times t}$. We then find the highest-similar time step pair, $\text{argmax}_{i,j}(M)$. We hypothesize that the highest-similar time step pair in $M$ yields a strong point to transition between the query and match as the audio spectra are most aligned. 

After finding the transition point, we perform a crossfade to further blend together the query and match audio clip. However, as previously mentioned, certain types of audio may benefit from different length crossfades \cite{lupsa2020audio}. Figure \ref{fig:examplesim} shows two examples of audio matches that require different crossfades. When matching the strikes of a hammer and knife, long-duration crossfades result in the impacts overlapping eachother and resulting in a blurry, low-quality transition. When matching a blender and motorcycle, the audio exhibits more noise throughout the sample, that benefits from longer crossfades as both noisy samples blend into eachother slowly. 

We model this characteristic based on the \textit{variance} of the spectrogram similarity matrix based on the hypothesis that audio pairs that exhibit high variance in their similarity matrix (e.g. impulsive sounds) require little-to-no crossfading, while audio pairs that exhibit low variance in their similarity matrix (e.g. noisy, static sounds) benefit from longer crossfades, as they have plausible transition points across multiple time steps. We use the inverse-variance of the computed pair similarity matrix to adaptively determine crossfade length, named ``Adaptive Crossfade":




\vspace{-1em}

\begin{equation}
l_{\text{crossfade}} = \frac{1}{Var(\overline{M}) \phi};  \overline{M} = \frac{S_{Q_i}^T S_{M_j}}{||S_{Q_i}||||S_{M_j}||} \forall i,j \in \{1,...,t\}
\end{equation}

\begin{table*}[ht!]
\setlength{\tabcolsep}{0.8em}
\centering
\begin{tabular}{lccccccc}
\hline
Retrieval Methods&Dataset&R-mAP&HR@1&HR@2&HR@5&P@5&P@10 \\ \hline
Random  & \multirow{6}{*}{AudioSet \cite{audioset}}   &.1093&.0392&.1373&.3235&.0804&.0794 \\
                MFCC             &  &.4111&.3725&.5196&.6961&.3510&.3206  \\ 
Mel-Spectrogram                      &  &.3318&.3529&.5392&.6569&.3157&.2882  \\
ImageBind \cite{imagebind}       &  &.5623&.6471&.7745&.9314&.5137&.4696  \\
CLAP \cite{clap}                 &  &.7225&.7843&.9314&.9608&.6765&.5990  \\ 
\textbf{Split-and-Contrast (Ours)}            &  &\textbf{.7656}&\textbf{.8333}&\textbf{.9608}&\textbf{.9804}&\textbf{.7216}&\textbf{.6069}  \\ \hline

Random   & \multirow{6}{*}{MovieClips} &.1176&.1061&.1364&.3030&.0727&.0742  \\
MFCC                             &  &.3266&.3636&.5000&.6667&.2576&.2197  \\
Mel-Spectrogram                      &  &.3337&.3485&.5606&.7273&.3758&.3258  \\
ImageBind \cite{imagebind}       &  &.5209&.4697&.6212&.7576&.4939&.4955 \\
CLAP \cite{clap}                 &  &.7729&.7424&.8939&.9848&.7636&.7136  \\ 
\textbf{Split-and-Contrast (Ours)}          &  &\textbf{.7995}&\textbf{.8788}&\textbf{.9394}&\textbf{1.000}&\textbf{.7758}&\textbf{.7227}  \\ \hline
\end{tabular}
\vspace{-0.5em}
\caption{Audio retrieval results on the labeled audio match cut evaluation set from AudioSet \cite{audioset} and MovieClips.}
\label{tab:retrieval}
\vspace{-1.5em}
\end{table*} 
\noindent Here, $\phi$ controls the scaling of the relationship of the similarity matrix variance to the crossfade length. We use a value of $\phi=8$. For the crossfade, we use a square-root window for fade-in and fade-out, with length and overlap of $l_{\text{crossfade}}$ seconds. We use the same Mel-Spectrogram parameters described in Section \ref{ssec:datacollection}. Note we use dot product to find the most similar time step pair and use cosine similarity in calculating the matrix variance to keep values bounded in a defined range. We found that using dot product takes the spectrogram magnitude into account (via un-normalized features) and as a result the transition point often occurs on time steps with higher energies, like strikes and impacts rather than quiet portions, which aligns well with many real-world audio match cuts.






\begin{table}[t]
\vspace{0.5em}
\setlength{\tabcolsep}{0.2em}
\centering
\begin{tabular}{lc}
\hline
Transition Methods                            & Transition Score (0-3) \\ \hline
Concatenation                      &     0.821          \\
Crossfade (0.25s)            &    1.750           \\
Crossfade (0.5s)            &     1.714        \\
Max-Sub-Spectrogram (Max-SS) (Ours)                     &     1.107          \\
\textbf{Max-SS + Adaptive Crossfade (Ours)} &    \textbf{2.143}           \\ \hline
\end{tabular}
\vspace{-0.5em}
\caption{Transition scores on audio transition methods.}
\label{tab:transition}
\vspace{-1.5em}
\end{table}

\vspace{-0.5em}

\section{Experiments}
\label{sec:experiments}

\subsection{Evaluation Metrics}
\label{ssec:evalmetrics}

To evaluate audio retrieval performance, we use multiple standard metrics that are widely used across various retrieval tasks. Specifically, we measure retrieval mean average precision (R-mAP), $\text{hit-rate}@K$, and $\text{precision}@K$ metrics. These metrics align well with the real-world use case of our proposed framework, where an editor is provided $K$ audio match cuts to choose from, with the goal of the $K$ recommendations being high-quality audio match cuts.

For evaluating transition quality, we construct criteria to grade the overall quality of the audio transition of a positive audio match pair. We create four criteria, ranging from $0-3$ of increasing transition quality: 0) Transition is poor and directly noticeable. 1) Transition is noticeable but is still a fluid transition. 2) Transition is high-quality that strongly matches either rhythm or timbre/pitch. 3) Transition is imperceptible, the transition point cannot be directly heard. 

\vspace{-0.5em}
\subsection{Retrieval Evaluation}
\label{ssec:retrievaleval}

Table \ref{tab:retrieval} shows qualitative audio match cut retrieval performance of multiple baseline methods against our proposed method. As shown, both the MFCC and Mel-Spectrogram representations are able to outperform random selection of audio matches, showing that simple non-learnable representations are able to effectively retrieve audio match cut candidates. However, when comparing large-scale deep audio representations ImageBind \cite{imagebind} and CLAP \cite{clap}, we see that they significantly outperform the non-learnable representations, with CLAP outperforming ImageBind \cite{imagebind} across all metrics. Although models like CLAP are trained for other tasks like language-audio alignment, the learned representations still are effective in audio-to-audio retrieval as the highly-similar samples are often also perceptually similar, the main criteria for creating audio match cuts. Finally, we see that our ``Split-and-Contrast" scheme outperforms CLAP \cite{clap} and all other methods across all retrieval metrics, showing our self-supervised objective is effective for better aligning audio representations for the audio match cut task.  



\vspace{-0.5em}
\subsection{Transition Evaluation}
\label{ssec:transitioneval}

To evaluate the quality of transition methods once an audio match is retrieved, we score the transition quality of 27 Audioset and 41 Movieclips positive matches. Table \ref{tab:transition} shows the average transition scores for multiple baseline transition methods and our proposed method, with and without crossfading. Simple concatenation of the query and match audio often results in artifacts and audible discontinuities, which degrade the transition quality as the exact borders of audio may not be perfectly aligned. When performing crossfading at multiple time lengths, significantly higher-quality match cuts are produced as discontinuties and slight differences in spectra are blended away with the crossfade. When comparing our method of selecting a specific transition point, named ``Max-SS", we see that it outperforms concatenation, showing that selecting a more optimal transition point within the 1-second query and match often results in a higher quality transition. When adding our proposed adaptive crossfading, we see the best transition performance, showing that the addition of selecting the optimal transition point and adaptively fading based on the audio characteristics outperforms each baseline. 

We highlight that the performance of the transition methods are often a function of how perceptually-similar the retrieved audio match is. The more the query and retrieved audio match, the less the need for advanced transition methods as they already transition from one another well. For very high-quality match retrievals, simple transitions may result in audio match cuts of similar perceptual quality to our proposed method. However, our method allows for the alignment of the cut on specific sound events, like impacts and strikes. Therefore, the specific transition method is left for user choice, depending on the type of audio match cut that is desired. 



\vspace{-0.5em}
\section{Conclusion}
\label{sec:conclusion}

In this paper, we introduce a framework to automatically find and create audio match cuts, an advanced video editing technique used in videos and movies. Analogous to visual match cutting \cite{matchcut}, this work can be used to aid in the automatic creation of trailers, edits, montages, and other videos by creating high-quality audio match cuts that are interesting and appealing to viewers. In the future, we hope to explore more advanced audio blending methods beyond crossfading, in addition creating audio-visual match cuts by incorporating the visual modality, with the ability to control specific audio-visual characteristics of the desired match cut.

\clearpage




\bibliographystyle{IEEEbib}
\bibliography{strings,refs}

\begin{thebibliography}{10}

\bibitem{cutting2016evolution}
James~E Cutting,
\newblock ``The evolution of pace in popular movies,''
\newblock {\em Cognitive research: principles and implications}, vol. 1, pp.
  1--21, 2016.

\bibitem{kozlovic2007anatomy}
Anton~Karl Kozlovic,
\newblock ``Anatomy of film,''
\newblock {\em Kinema: A Journal for Film and Audiovisual Media}, vol. 1, 2007.

\bibitem{douglass1996art}
John~S Douglass and Glenn~P Harnden,
\newblock ``The art of technique: An aesthetic approach to film and video
  production,''
\newblock 1996.

\bibitem{thompson2013grammar}
Roy Thompson and Christopher~J Bowen,
\newblock {\em Grammar of the Edit},
\newblock Taylor \& Francis, 2013.

\bibitem{matchcut}
Boris Chen, Amir Ziai, Rebecca~S Tucker, and Yuchen Xie,
\newblock ``Match cutting: Finding cuts with smooth visual transitions,''
\newblock in {\em Proceedings of the IEEE/CVF Winter Conference on Applications
  of Computer Vision}, 2023, pp. 2115--2125.

\bibitem{irie2010automatic}
Go~Irie, Takashi Satou, Akira Kojima, Toshihiko Yamasaki, and Kiyoharu Aizawa,
\newblock ``Automatic trailer generation,''
\newblock in {\em Proceedings of the 18th ACM international conference on
  Multimedia}, 2010, pp. 839--842.

\bibitem{ave}
Dawit~Mureja Argaw, Fabian~Caba Heilbron, Joon-Young Lee, Markus Woodson, and
  In~So Kweon,
\newblock ``The anatomy of video editing: a dataset and benchmark suite for
  ai-assisted video editing,''
\newblock in {\em European Conference on Computer Vision}. Springer, 2022, pp.
  201--218.

\bibitem{moviecuts}
Alejandro Pardo, Fabian~Caba Heilbron, Juan~Le{\'o}n Alc{\'a}zar, Ali Thabet,
  and Bernard Ghanem,
\newblock ``Moviecuts: A new dataset and benchmark for cut type recognition,''
\newblock in {\em European Conference on Computer Vision}. Springer, 2022, pp.
  668--685.

\bibitem{pardo2021learning}
Alejandro Pardo, Fabian Caba, Juan~Le{\'o}n Alc{\'a}zar, Ali~K Thabet, and
  Bernard Ghanem,
\newblock ``Learning to cut by watching movies,''
\newblock in {\em Proceedings of the IEEE/CVF International Conference on
  Computer Vision}, 2021, pp. 6858--6868.

\bibitem{shen2022autotransition}
Yaojie Shen, Libo Zhang, Kai Xu, and Xiaojie Jin,
\newblock ``Autotransition: Learning to recommend video transition effects,''
\newblock in {\em European Conference on Computer Vision}. Springer, 2022, pp.
  285--300.

\bibitem{pei2023automatch}
Sen Pei, Jingya Yu, Qi~Chen, and Wozhou He,
\newblock ``Automatch: A large-scale audio beat matching benchmark for boosting
  deep learning assistant video editing,''
\newblock {\em arXiv preprint arXiv:2303.01884}, 2023.

\bibitem{spaceodyssey}
Stanley Kubrick and Arthur~C. Clarke,
\newblock ``2001: A space odyssey,'' 1968.

\bibitem{narnia}
Andrew Adamson,
\newblock ``The chronicles of narnia: The lion, the witch and the wardrobe,''
  2005.

\bibitem{six2014panako}
Joren Six and Marc Leman,
\newblock ``Panako: a scalable acoustic fingerprinting system handling
  time-scale and pitch modification,''
\newblock in {\em 15th International society for music information retrieval
  conference (ISMIR-2014)}, 2014.

\bibitem{fenet2011scalable}
S{\'e}bastien Fenet, Ga{\"e}l Richard, Yves Grenier, et~al.,
\newblock ``A scalable audio fingerprint method with robustness to
  pitch-shifting.,''
\newblock in {\em ISMIR}, 2011, pp. 121--126.

\bibitem{banerjee2022wav2tok}
Adhiraj Banerjee and Vipul Arora,
\newblock ``wav2tok: Deep sequence tokenizer for audio retrieval,''
\newblock in {\em The Eleventh International Conference on Learning
  Representations}, 2022.

\bibitem{chang2021neural}
Sungkyun Chang, Donmoon Lee, Jeongsoo Park, Hyungui Lim, Kyogu Lee, Karam Ko,
  and Yoonchang Han,
\newblock ``Neural audio fingerprint for high-specific audio retrieval based on
  contrastive learning,''
\newblock in {\em ICASSP 2021-2021 IEEE International Conference on Acoustics,
  Speech and Signal Processing (ICASSP)}. IEEE, 2021, pp. 3025--3029.

\bibitem{vande2018raw}
Len Vande~Veire and Tijl De~Bie,
\newblock ``From raw audio to a seamless mix: creating an automated dj system
  for drum and bass,''
\newblock {\em EURASIP Journal on Audio, Speech, and Music Processing}, vol.
  2018, no. 1, pp. 1--21, 2018.

\bibitem{chen2022automatic}
Bo-Yu Chen, Wei-Han Hsu, Wei-Hsiang Liao, Marco A~Mart{\'\i}nez Ram{\'\i}rez,
  Yuki Mitsufuji, and Yi-Hsuan Yang,
\newblock ``Automatic dj transitions with differentiable audio effects and
  generative adversarial networks,''
\newblock in {\em ICASSP 2022-2022 IEEE International Conference on Acoustics,
  Speech and Signal Processing (ICASSP)}. IEEE, 2022, pp. 466--470.

\bibitem{audioset}
Jort~F Gemmeke, Daniel~PW Ellis, Dylan Freedman, Aren Jansen, Wade Lawrence,
  R~Channing Moore, Manoj Plakal, and Marvin Ritter,
\newblock ``Audio set: An ontology and human-labeled dataset for audio
  events,''
\newblock in {\em 2017 IEEE international conference on acoustics, speech and
  signal processing (ICASSP)}. IEEE, 2017, pp. 776--780.

\bibitem{clap}
Yusong Wu, Ke~Chen, Tianyu Zhang, Yuchen Hui, Taylor Berg-Kirkpatrick, and
  Shlomo Dubnov,
\newblock ``Large-scale contrastive language-audio pretraining with feature
  fusion and keyword-to-caption augmentation,''
\newblock in {\em ICASSP 2023-2023 IEEE International Conference on Acoustics,
  Speech and Signal Processing (ICASSP)}. IEEE, 2023, pp. 1--5.

\bibitem{imagebind}
Rohit Girdhar, Alaaeldin El-Nouby, Zhuang Liu, Mannat Singh, Kalyan~Vasudev
  Alwala, Armand Joulin, and Ishan Misra,
\newblock ``Imagebind: One embedding space to bind them all,''
\newblock in {\em Proceedings of the IEEE/CVF Conference on Computer Vision and
  Pattern Recognition}, 2023, pp. 15180--15190.

\bibitem{oord2018representation}
Aaron van~den Oord, Yazhe Li, and Oriol Vinyals,
\newblock ``Representation learning with contrastive predictive coding,''
\newblock {\em arXiv preprint arXiv:1807.03748}, 2018.

\bibitem{sohn2016improved}
Kihyuk Sohn,
\newblock ``Improved deep metric learning with multi-class n-pair loss
  objective,''
\newblock {\em Advances in neural information processing systems}, vol. 29,
  2016.

\bibitem{htsat}
Ke~Chen, Xingjian Du, Bilei Zhu, Zejun Ma, Taylor Berg-Kirkpatrick, and Shlomo
  Dubnov,
\newblock ``Hts-at: A hierarchical token-semantic audio transformer for sound
  classification and detection,''
\newblock in {\em ICASSP 2022-2022 IEEE International Conference on Acoustics,
  Speech and Signal Processing (ICASSP)}. IEEE, 2022, pp. 646--650.

\bibitem{adam}
Diederik~P Kingma and Jimmy Ba,
\newblock ``Adam: A method for stochastic optimization,''
\newblock {\em arXiv preprint arXiv:1412.6980}, 2014.

\bibitem{archibaldcross}
Fitzgerald~J Archibald,
\newblock ``Cross fade of digital audio streams,''
\newblock .

\bibitem{lupsa2020audio}
Lucian Lupsa-Tataru,
\newblock ``Audio fade-out profile shaping for interactive multimedia,''
\newblock 2020.

\end{thebibliography}

\end{document}